\documentclass[12pt]{article}
\def\text#1{\mbox{$#1$}}

\usepackage{amsfonts}
\begin{document}
\hfill{\hbox{hep-th/0102036}}
\begin{center}
{\LARGE \bf The $D_{n}$ Ruijsenaars-Schneider model \par}
\vskip 3em
{\large
\lineskip .75em
Kai Chen\footnote{E-mail: kai@phy.nwu.edu.cn},
Bo-yu Hou\footnote{E-mail: byhou@phy.nwu.edu.cn}\\
{\small \it Institute of Modern Physics, Northwest University, Xi'an
710069, China}}
\end{center}
\vskip 3em

\begin{abstract}
The Lax pair of the Ruijsenaars-Schneider model with
interaction potential of trigonometric type based on $D_{n}$ Lie algebra is
presented. We give a general form for the Lax pair and prove partial
results for small $n$. Liouville integrability of the corresponding system
follows a series of involutive Hamiltonians generated by the
characteristic polynomial of the Lax matrix. The rational case appears as a
natural degeneration and the nonrelativistic limit exactly leads to the
well-known Calogero-Moser system associated with $D_{n}$ Lie algebra.
\newline
\textbf{\noindent Keywords:} Lax pair; $D_{n}$ Lie algebra;
Ruijsenaars-Schneider(RS) model \newline
\textbf{\noindent PACS:} 02.20.+b, 11.10.Lm, 03.80.+r
\end{abstract}

\newpage

\section{Introduction}

\setcounter{equation}{0}

Since a relativistic version of the
Calogero-Moser(CM) model was first introduced by Ruijsenaars and Schneider%
\cite{r3, r1, r2}, much interest have been focused on this model and its
nonrelativistic counterpart. It is completely integrable many-body
Hamiltonian system describing a one-dimensional $n$-particle system with
pairwise interaction. The study has lead to fascinating mathematics and
application from lattice models in statistics physics\cite{h1, nksr}, to the
field theory and gauge theory \cite{n}. e.g. to the Seiberg-Witten theory%
\cite{bm1} etc. For a review see \cite{hp2, Marbook, g00}, and references
therein.

Recently, the Lax pairs for the CM models in various root system have been
constructed by Olshanetsky and Perelomov\cite{op} using reduction on
symmetric space, further given by Inozemtsev in \cite{in}. Afterwards,
D'Hoker and Phong\cite{hp1} succeeded in constructing the Lax pairs with
spectral parameter for each of the finite dimensional Lie algebra, as well
as the introduction of untwisted and twisted Calogero-Moser systems. Bordner
\textit{et al}\cite{bcs,bcs2,bcs3} give two types universal realization for
the Lax pairs associated to all of the Lie algebra: the root type and the
minimal type, with and without spectral parameters. Even for all of the
Coxeter group, the construction has been obtained in \cite{bcs1}. All of
them do not apply the reduction method for under which condition one will
confront some obstruction\cite{hm} but using pure Lie algebra construction.
In \cite{hm}, Hurtubise and Markman utilize so called ``structure group'',
which combines semi-simple group and Weyl group, to construct the CM systems
associate with Hitchin system, which in some degree generalizes the results
of Refs. \cite{hp1,bcs,bcs2,bcs3,bcs1}. Furthermore, the quantum version of
the generalization have been developed in \cite{bms,kps} at least for
degenerate potentials of trigonometric after the works of Olshanetsky and
Perelomov\cite{op1}.

So far as for the RS model, only the Lax pair of the $A_{N-1}$ type RS model
was obtained \cite{r1,nksr,bc,kz,s1,s2} and succeeded in recovering it by
applying Hamiltonian reduction procedure on two-dimensional current group
\cite{aru}. Although the commutative operators for RS model based on various
type Lie algebra have been given by Komori and co-workers \cite{ko1}, Diejen%
\cite{di,di1} and Hasegawa \textit{et al}\cite{h1,h2}, the Lax integrability
(or the Lax representation) of the other type RS model is still an open
problem\cite{bm1}.

In recent work of Refs. \cite{kai3} and \cite{Che00}, we have succeeded in
constructing the Lax pairs for the rational and trigonometric $C_{n}$ and $BC_{n}
$ RS systems. Following that, the $r$-matrix structure for them have been
derived by Avan \textit{et al} in \cite{Avan}. Moreover we give the more
general elliptic $C_{n}$ and $BC_{n}$ RS systems in \cite{Che01} and
calculate their spectral curves. In this paper, we concentrate on
generalizing the construction to the $D_{n}-$type trigonometric
Ruijsenaars-Schneider model. It turns out that there surely exists a Lax
pair for this system. By revealing the symmetry property of this model, we
shall give a general form of the Lax pair for generic  $n$ and verify its
rationality at least for small $n$ such as $n=2,3,4,5,6$. Its integrability
in Liouville sense is also depicted by giving $n$ involutive integrals of
motion. We also perform its non-relativistic limit that coincides exactly
with the previous known result for the $D_{n}$ Calogero-Moser system. The
rational degeneration of this system is also remarked.

The paper is organized as follows. The basic materials of the $D_{n}$ RS
model are introduced in Section \ref{model}, where we propose a
self-consistent dynamical system associating with the root system of $D_{n}$
. This includes construction of Hamiltonian for the $D_{n}$ RS system
together with its symmetry analysis etc. The main results are showed in
Sections \ref{laxpair}. In Section \ref{laxpair}, we present a Lax pair and
obtain an  explicit general form for the Lax pair by imposing some
additional symmetry constraints. Section \ref{limit} is devoted to deriving
the non-relativistic counterpart, the Calogero-Moser model. Following is
some remarks for the degenerate limit of rational case. We conclude with some
remarks on our constructions in the last section.

\section{Model and equations of motion}

\setcounter{equation}{0} \label{model}

Let us first review the basic materials about the $D_{n}$ RS model. Though
much progress have been made for generalization about the RS model\cite
{r2,ko1,di,kai3,Che00,Che01}, there is no any result for the system which
associates with the root system of $D_{n}$. Even up to now we do not know
how to define its Hamiltonian. But now we will give a reasonable definition
for this system which will be seen later.

In terms of the canonical variables $p_{i}$, $x_{i}(i,j=1,\ldots ,n)$
enjoying in the canonical Poisson bracket
\begin{equation}
\{p_{i},p_{j}\}=\{x_{i},x_{j}\}=0,\mbox{$ \ \ \ \ \ \ \ \ \ \ \ \ \ \ $}%
\{x_{i},p_{j}\}=\delta _{ij},  \label{poisson}
\end{equation}
we give firstly the Hamiltonian of $D_{n}$ RS system
\begin{equation}
H=\sum_{i=1}^{n}\Big(e^{p_{i}}\,\prod_{k\neq
i}^{n}(f(x_{ik})f(x_{i}+x_{k}))+e^{-p_{i}}\,\prod_{k\neq
i}^{n}(g(x_{ik})g(x_{i}+x_{k}))\Big),  \label{hami}
\end{equation}
where
\begin{eqnarray}
f(x) &:&=\frac{\sin (x-\gamma )}{\sin (x)},  \nonumber \\
g(x) &:&=f(x)|_{\gamma \rightarrow -\gamma
},~~~~~~~~~~~~~x_{ik}:=x_{i}-x_{k},
\end{eqnarray}
and $\gamma $ denotes the coupling constant. Then the canonical equations of
motion could be
\begin{eqnarray}
\dot{x_{i}} &=&\{x_{i},H\}=e^{p_{i}}b_{i}-e^{-p_{i}}b_{i}^{^{\prime }},
\label{equ1} \\
\dot{p_{i}} &=&\{p_{i},H\}=\sum_{j\neq i}^{n}\Big(e^{p_{j}}b_{j}\big(%
h(x_{ji})-h(x_{j}+x_{i})\big)  \nonumber \\
&&+e^{-p_{j}}b_{j}^{^{\prime }}\big(\hat{h}(x_{ji})-\hat{h}(x_{j}+x_{i})\big)%
\Big)  \nonumber \\
&&-e^{p_{i}}b_{i}\Big(\sum_{j\neq i}^{n}\big(h(x_{ij})+h(x_{i}+x_{j})\big)%
\Big)  \nonumber \\
&&-e^{-p_{i}}b_{i}^{^{\prime }}\Big(\sum_{j\neq i}^{n}\big(\hat{h}(x_{ij})+%
\hat{h}(x_{i}+x_{j})\big)\Big),  \label{equ2}
\end{eqnarray}
where
\begin{eqnarray}
h(x):= &&\frac{d\ln f(x)}{dx},\mbox{$ \ \ \ \ \ \ \ \ \ \ \ $}\hat{h}(x):=%
\frac{d\ln g(x)}{dx},  \nonumber \\
b_{i} &=&\,\prod_{k\neq i}^{n}\Big(f(x_{i}-x_{k})f(x_{i}+x_{k})\Big),
\nonumber \\
b_{i}^{^{\prime }} &=&\,\prod_{k\neq i}^{n}\Big(g(x_{i}-x_{k})g(x_{i}+x_{k})%
\Big).
\end{eqnarray}
Here,\vspace{1pt} of course $x_{i}=x_{i}(t)$, $p_{i}=p_{i}(t)$ and the
superimposed dot denotes $t$-differentiation.

For the convenience of analysis of symmetry, let us first give vector
representation of $D_{n}$ Lie algebra. Introducing an $n-$dimensional
orthonormal basis of ${\mathbb R}^{n}$,
\begin{equation}
e_{j}\cdot e_{k}=\delta _{j,k},\quad j,k=1,2,\ldots ,n,
\end{equation}
then the sets of roots $\Delta $ and vector weights $\Lambda $ of $D_{n}$
are:
\begin{eqnarray}
\Delta &=&\{\pm (e_{j}-e_{k}),\pm (e_{j}+e_{k}):\mbox{$ \ $}j,k=1,2,\ldots
,n \mbox{ and }j<k\},\quad  \label{anroot} \\
\Lambda &=&\{e_{j},-e_{j}:\mbox{$ \ $}j=1,2,\ldots ,n\mbox{$\ $}\}.
\label{anwei}
\end{eqnarray}
The dynamical variables are canonical coordinates $\{x_{j}\}$ and their
canonical conjugate momenta $\{p_{j}\}$ with the Poisson brackets of Eq.(
\ref{poisson}). In a general sense, we denote them by $n$ dimensional
vectors $x$ and $p$,
\[
x=(x_{1},x_{2},\ldots ,x_{n})\in {\mathbb R}^{n},\quad p=(p_{1},p_{2},\ldots
,p_{n})\in {\mathbb R}^{n}.
\]
So that the scalar products of $x$ and $p$ with the roots $\alpha \cdot x$, $%
p\cdot \beta $, etc. can be defined. The Hamiltonian of Eq.(\ref{hami}) can
be rewritten as
\begin{equation}
H=\frac{1}{2}\sum_{\mu \in \Lambda }\left( \exp \left( \mu \cdot p\right)
\,\prod_{\Delta \ni \beta =\mu -\nu }f(\beta \cdot x){+}\exp \left( -\mu
\cdot p\right) \,\prod_{\Delta \ni \beta =-\mu +\nu }g(\beta \cdot x)\right)
,  \label{ham}
\end{equation}
Here, the condition $\Delta \ni \beta =\mu -\nu $ means that the summation
is over roots $\beta $ such that for $\exists \nu \in \Lambda $
\begin{equation}
\mu -\nu =\beta \in \Delta .
\end{equation}
So does for $\Delta \ni \beta =-\mu +\nu $.

From the above-mentioned data, we can see that the definition for the
Hamiltonian is reasonable and well-defined whose form Eq.(\ref{hami}) or
Eq.(\ref{ham}) is similar to the one given in \cite{kai3,Che00,Che01}.

\section{Construction of the Lax pair}

\setcounter{equation}{0} \label{laxpair}

In this section, we concentrate our treatment to the explicit form of the
Lax pair for the $D_{n}$ RS system. Therefore, some previous results, as
well as new results, could now be obtained in a more straightforward manner
by using the Lax pair.

\subsection{Derivation of the Lax matrix for the $D_{n}$ RS model}

Similar to the definitions of the Lax matrixes for the $C_{n}$ and $BC_{n}$
RS models given in \cite{Che00}, we suppose the Lax matrix for the $D_{n}$
RS model is one  $2n\times 2n$ matrix as follows:

\begin{equation}
L=\left(
\begin{array}{ll}
A & B \\
C & D
\end{array}
\right) ,  \label{lmat}
\end{equation}
where $A$, $B$, $C$, $D$ are $n\times n$ matrixes(hereafter, we use the
indices $i,j=1,\ldots ,n$)

\begin{eqnarray}
A_{ij} &=&e^{p_{j}}b_{j}\frac{\sin \gamma }{\sin (x_{ij}+\gamma )},\ \ \ \ \
\ \ \ \ \ \ \ D_{ij}=e^{-p_{j}}b_{j}^{^{\prime }}\frac{\sin \gamma }{\sin
(x_{ji}+\gamma )},  \nonumber \\
B_{ij} &=&(1-\delta _{ij})e^{-p_{j}}b_{j}^{^{\prime }}\frac{\sin \gamma }{
\sin (x_{i}+x_{j}+\gamma )}+\delta _{ij}e^{-p_{i}}\frac{b_{i}^{^{\prime }}}{
w_{i}}\widetilde{B}_{ii},  \nonumber \\
C_{ij} &=&(1-\delta _{ij})e^{p_{j}}b_{j}\frac{\sin \gamma }{\sin
(-x_{i}-x_{j}+\gamma )}+\delta _{ij}e^{p_{i}}\frac{b_{i}}{w_{i}^{^{\prime }}}
\widetilde{C}_{ii}.  \label{l1}
\end{eqnarray}
Here, the notations of $w_{i},w_{i}^{^{\prime }},v_{i}$ are

\begin{eqnarray}
w_{i} &:&=\prod_{j\neq i}^{n}\sin (x_{i}+x_{j}+\gamma )\sin (x_{ij}+\gamma ),
\nonumber \\
w_{i}^{^{\prime }} &:&=\prod_{j\neq i}^{n}\sin (x_{i}+x_{j}-\gamma )\sin
(x_{ij}-\gamma ),  \nonumber \\
v_{i} &:&=\prod_{j\neq i}^{n}\sin (x_{i}+x_{j})\sin (x_{ij}),
\end{eqnarray}
and $\widetilde{B}_{ii}$, $\widetilde{C}_{ii}$, the diagonal part of block
matrixes $B$ and $C$, are unknown and have to be solved later.

In order to obtain the explicit form of $\widetilde{B}_{ii}$, $\widetilde{C }
_{ii}$, we also assume the inverse of $L$ the following $2n\times 2n$
matrix(similar to the form for the $C_{n}$ and $BC_{n}$ cases)

\begin{equation}
L^{-1}=\left(
\begin{array}{ll}
\widehat{A} & \widehat{B} \\
\widehat{C} & \widehat{D}
\end{array}
\right) ,
\end{equation}
where $\widehat{A}$, $\widehat{B}$, $\widehat{C}$, $\widehat{D}$ are $%
n\times n$ matrixes

\begin{eqnarray}
\widehat{A}_{ij} &=&e^{-p_{i}}b_{j}^{^{\prime }}\frac{-\sin \gamma }{\sin
(x_{ij}-\gamma )},\ \ \ \ \ \ \ \ \ \ \ \ \widehat{D}_{ij}=e^{p_{i}}b_{j}
\frac{-\sin \gamma }{\sin (x_{ji}-\gamma )},  \nonumber \\
\widehat{B}_{ij} &=&(1-\delta _{ij})e^{-p_{i}}b_{j}\frac{-\sin \gamma }{\sin
(x_{i}+x_{j}-\gamma )}+\delta _{ij}e^{-p_{i}}\frac{b_{i}}{w_{i}^{^{\prime }}}
\widetilde{C}_{ii},  \nonumber \\
\widehat{C}_{ij} &=&(1-\delta _{ij})e^{p_{i}}b_{j}^{^{\prime }}\frac{-\sin
\gamma }{\sin (-x_{i}-x_{j}-\gamma )}+\delta _{ij}e^{p_{i}}\frac{
b_{i}^{^{\prime }}}{w_{i}}\widetilde{B}_{ii}.
\end{eqnarray}
If we impose an additional condition for $\widetilde{B}_{ii}$ and $%
\widetilde{C}_{ii}$ as following

\vspace{1pt}
\begin{equation}
\widetilde{C}_{ii}=\widetilde{B}_{ii}|_{\gamma \rightarrow -\gamma },
\label{reflection}
\end{equation}
then the equation

\begin{equation}
L\cdot L^{-1}=Id  \label{id}
\end{equation}
can be solved and the solution reads

\begin{eqnarray}
\widetilde{B}_{ii} &=&\frac{w_{i}}{b_{i}^{^{\prime }}}\Big( %
1-b_{i}b_{i}^{^{\prime }}-\sin ^{2}\gamma \sum_{j\neq i}^{n}(\frac{
b_{i}^{^{\prime }}b_{k}}{\sin ^{2}(x_{ik}+\gamma )}+\frac{b_{i}^{^{\prime
}}b_{k}^{^{\prime }}}{\sin ^{2}(x_{i}+x_{k}+\gamma )})\Big)^{\frac{1}{2}}
\nonumber \\
\widetilde{C}_{ii} &=&\widetilde{B}_{ii}|_{\gamma \rightarrow -\gamma }.
\label{lmat1}
\end{eqnarray}
So that

\begin{eqnarray}
B_{ii} &=&e^{-p_{i}}\frac{b_{i}^{^{\prime }}}{w_{i}}\widetilde{B}_{ii},
\nonumber \\
C_{ii} &=&e^{p_{i}}\frac{b_{i}}{w_{i}^{^{\prime }}}\widetilde{C}
_{ii}=B_{ii}|_{\gamma \rightarrow -\gamma ,p_{i}\rightarrow -p_{i}},
\label{lmat2}
\end{eqnarray}

\vspace{0.3cm} \noindent\textbf{Remarks:}

\begin{quotation}
The above solution of Eq.(\ref{lmat1}) and Eq.(\ref{lmat2}) is obtained only
by diagonal part of Eq.(\ref{id}). It is not easy to verify if the
off-diagonal part is consistent to the diagonal part due to the complicated
functional relations. But for small $n$ such as $n=2,3,4,5,6$ we can surely
check it is the very unique solution. In addition, it is unfortunate that we
are not able to give more simple forms for $B_{ii}$ and $C_{ii}$. Here only
for $n=2,3,4$ we work out the following results to shed a light on its
appearance:
\end{quotation}

\begin{itemize}
\item  for $n=2$
\begin{eqnarray}
\widetilde{B}_{ii} &=&\sin ^{2}\gamma ,  \nonumber \\
\widetilde{C}_{ii} &=&\sin ^{2}\gamma ,
\end{eqnarray}

\item  for $n=3$
\begin{eqnarray}
\widetilde{B}_{ii} &=&\frac{1}{2}\sin ^{2}\gamma \Big(w_{i}\sum_{j\neq i}^{n}%
\frac{1}{\sin (x_{i}+x_{j}+\gamma )\sin (x_{ij}+\gamma )}  \nonumber \\
&&+v_{i}\sum_{j\neq i}^{n}\frac{1}{\sin (x_{i}+x_{j})\sin (x_{ij})}\Big),
\nonumber \\
\widetilde{C}_{ii} &=&\frac{1}{2}\sin ^{2}\gamma \Big(w_{i}^{^{\prime
}}\sum_{j\neq i}^{n}\frac{1}{\sin (x_{i}+x_{j}-\gamma )\sin (x_{ij}-\gamma )}
\nonumber \\
&&+v_{i}\sum_{j\neq i}^{n}\frac{1}{\sin (x_{i}+x_{j})\sin (x_{ij})}\Big)
\nonumber \\
&=&\widetilde{B}_{ii}|_{\gamma \mapsto -\gamma },
\end{eqnarray}

\item  for $n=4$
\begin{eqnarray}
\widetilde{B}_{ii} &=&\frac{1}{2}\sin ^{2}\gamma \Big(w_{i}\sum_{j\neq i}^{n}%
\frac{1}{\sin (x_{i}+x_{j}+\gamma )\sin (x_{ij}+\gamma )}  \nonumber \\
&&+v_{i}\sum_{j\neq i}^{n}\frac{1}{\sin (x_{i}+x_{j})\sin (x_{ij})}-\sin
^{2}\gamma \sin ^{2}(2x_{i}+\gamma )\Big),  \nonumber \\
\widetilde{C}_{ii} &=&\frac{1}{2}\sin ^{2}\gamma \Big(w_{i}^{^{\prime
}}\sum_{j\neq i}^{n}\frac{1}{\sin (x_{i}+x_{j}-\gamma )\sin (x_{ij}-\gamma )}
\nonumber \\
&&+v_{i}\sum_{j\neq i}^{n}\frac{1}{\sin (x_{i}+x_{j})\sin (x_{ij})}-\sin
^{2}\gamma \sin ^{2}(2x_{i}-\gamma )\Big)  \nonumber \\
&=&\widetilde{B}_{ii}|_{\gamma \mapsto -\gamma }.
\end{eqnarray}

With this Lax matrix $L$ of Eq.(\ref{lmat}), we could rewrite the
Hamiltonian as
\end{itemize}

\begin{equation}
H=\sum_{j=1}^{n}(e^{p_{j}}b_{j}+e^{-p_{j}}b_{j}^{^{\prime }})=trL.
\end{equation}
The involutive $n$ Hamiltonians can be generated by the characteristic
polynomial of the Lax matrix
\begin{equation}
\det (L-v\cdot
Id)=\sum_{j=0}^{2n}(-1)^{j}(v^{j}+v^{2n-j})H_{j}+(-v)^{n}H_{n},
\end{equation}
with
\begin{equation}
\{H_{i},H_{j}\}=0,\mbox{$ \ \ \ \ \ $}i,j=1,2,\ldots ,n.\mbox{$\ \ \ \ $}
\end{equation}
e.g. for $n=2$,
\begin{equation}
\det (L-v\cdot Id)=v^{4}-H_{1}v^{3}+H_{2}v^{2}-H_{1}v+1,
\end{equation}
the function-independent Hamiltonian flows $H$ and $H_{2}$ are
\begin{eqnarray}
H_{1} &=&H=e^{p_{1}}f(x_{12})f(x_{1}+x_{2})+e^{-p_{1}}g(x_{12})g(x_{1}+x_{2})
\nonumber \\
&&+e^{p_{2}}f(x_{21})f(x_{2}+x_{1})+e^{-p_{2}}g(x_{21})g(x_{2}+x_{1}), \\
H_{2} &=&2\Big(f(x_{12})g(x_{12})+f(x_{1}+x_{2})g(x_{1}+x_{2})\Big)
\nonumber \\
&&e^{p_{1}+p_{2}}f(x_{1}+x_{2})^{2}+e^{-p_{1}-p_{2}}g(x_{1}+x_{2})^{2}
\nonumber \\
&&+e^{p_{1}-p_{2}}f(x_{12})^{2}+e^{p_{2}-p_{1}}g(x_{12})^{2}+const,
\end{eqnarray}
where $const=-2$. For $n=3,$ we have
\begin{equation}
\det (L-v\cdot
Id)=v^{6}-H_{1}v^{5}+H_{2}v^{4}-H_{3}v^{3}+H_{2}v^{2}-H_{1}v^{1}+1,
\end{equation}
and

\begin{eqnarray}
H_{1} &=&H=\sum_{i=1}^{3}\Big(e^{p_{i}}\,\prod_{k\neq
i}^{3}f(x_{ik})f(x_{i}+x_{k})+e^{-p_{i}}\,\prod_{k\neq
i}^{3}g(x_{ik})g(x_{i}+x_{k})\,\Big),  \nonumber \\
H_{2} &=&\widetilde{H}_{2}-1, \\
H_{3} &=&\widetilde{H}_{3}-2H_{1},
\end{eqnarray}
here $\widetilde{H}_{2}$ and $\widetilde{H}_{3}$ are the involutive
Hamiltonians defined for the $D_{3}$ RS model by Diejen in \cite{di}
\begin{eqnarray}
H_{+}
&=&e^{(-p_{1}-p_{2}+p_{3})/2}f(-x_{1}-x_{2})f(-x_{1}+x_{3})f(-x_{2}+x_{3})
\nonumber \\
&&+e^{(-p_{1}+p_{2}-p_{3})/2}f(-x_{1}+x_{2})f(-x_{1}-x_{3})f(x_{2}-x_{3})
\nonumber \\
&&+e^{(p_{1}+p_{2}+p_{3})/2}f(x_{1}+x_{2})f(x_{1}+x_{3})f(x_{2}+x_{3})
\nonumber \\
&&+e^{(p_{1}-p_{2}-p_{3})/2}f(x_{12})f(x_{13})f(-x_{2}-x_{3}), \\
H_{-}
&=&e^{(-p_{1}-p_{2}-p_{3})/2}f(-x_{1}-x_{2})f(-x_{1}-x_{3})f(-x_{2}-x_{3})
\nonumber \\
&&+e^{(-p_{1}+p_{2}+p_{3})/2}f(-x_{1}+x_{2})f(-x_{1}+x_{3})f(x_{2}+x_{3})
\nonumber \\
&&+e^{(p_{1}-p_{2}+p_{3})/2}f(x_{12})f(x_{1}+x_{3})f(-x_{2}+x_{3})  \nonumber
\\
&&+e^{(p_{1}+p_{2}-p_{3})/2}f(x_{1}+x_{2})f(x_{13})f(x_{23}), \\
\widetilde{H}_{2} &=&H_{+}H_{-}, \\
\widetilde{H}_{3} &=&H_{+}^{2}+H_{-}^{2}.
\end{eqnarray}
We verify that these $H_{i}$ and $\widetilde{H}_{j}$ strictly Poisson
commute each other, which ensures the complete integrability of the $D_{2}$
and $D_{3}$ RS models (in Liouville sense).

\subsection{$M$ operator associating with $L$}

By comparing the symmetry of the $D_{n}$ RS model and $BC_{n}$ one, we
propose the following ansatz for $M$ operator associating with the Lax
matrix $L$ so that they satisfy

\begin{equation}
\dot{L}=\{L,H\}=\lbrack M,L\rbrack .  \label{laxequ}
\end{equation}
$M$ suppose to be another $2n\times 2n$ matrix with the form

\begin{eqnarray}
M=\left(
\begin{array}{ll}
\mathcal{A} & \mathcal{B} \\
\mathcal{C} & \mathcal{D}
\end{array}
\right),
\end{eqnarray}
where entries of $M$ are

\begin{eqnarray}
\mathcal{A}_{ij} &=&\cot (x_{ij})\Big(e^{p_{j}}b_{j}\frac{\sin \gamma }{\sin
(x_{ij}+\gamma )}+e^{-p_{i}}b_{j}^{^{\prime }}\frac{\sin \gamma }{\sin
(x_{ij}-\gamma )}\Big),~~~~~~~~~j\neq i,  \nonumber \\
\mathcal{D}_{ij} &=&\cot (x_{ji})\Big(e^{-p_{j}}b_{j}^{^{\prime }}\frac{\sin
\gamma }{\sin (x_{ji}+\gamma )}+e^{p_{i}}b_{j}\frac{\sin \gamma }{\sin
(x_{ji}-\gamma )}\Big),~~~~~~~~~j\neq i,  \nonumber \\
\mathcal{B}_{ij} &=&\cot (x_{i}+x_{j})\Big(e^{-p_{j}}b_{j}^{^{\prime }}\frac{%
\sin \gamma }{\sin (x_{i}+x_{j}+\gamma )}+e^{-p_{i}}b_{j}\frac{\sin \gamma }{%
\sin (x_{i}+x_{j}-\gamma )}\Big),~~~\ \ \ j\neq i,  \nonumber \\
\mathcal{C}_{ij} &=&\cot (-x_{i}-x_{j})\Big(e^{p_{j}}b_{j}\frac{\sin \gamma
}{\sin (-x_{i}-x_{j}+\gamma )}+e^{p_{i}}b_{j}^{^{\prime }}\frac{\sin \gamma
}{\sin (-x_{i}-x_{j}-\gamma )}\Big),~~~j\neq i,  \nonumber \\
\mathcal{A}_{ii} &=&-\Big(\sum_{j\neq i}^{n}\frac{\mathcal{A}_{ij}}{\cos
(x_{ij})}+\sum_{j=1}^{n}\frac{\mathcal{B}_{ij}}{\cos (x_{i}+x_{j})}\Big),
\nonumber \\
\mathcal{D}_{ii} &=&-\Big(\sum_{j\neq i}^{n}\frac{\mathcal{D}_{ij}}{\cos
(x_{ji})}+\sum_{j=1}^{n}\frac{\mathcal{C}_{ij}}{\cos (-x_{i}-x_{j})}\Big).
\label{me1}
\end{eqnarray}
If we impose $\mathcal{B}_{ii},\mathcal{C}_{ii}$ an additional symmetry
condition with

\begin{equation}
\mathcal{B}_{ii}=e^{2p_{i}}\mathcal{C}_{ii},
\end{equation}
verbose but straightforward calculations of equations

\begin{eqnarray}
\dot{L}_{ii} &=&\{L_{ii},H\}=(\lbrack M,L\rbrack )_{ii}  \nonumber \\
&=&\sum_{k\neq i}^{2n}(M_{ik}L_{ki}-L_{ik}M_{ki}),
\end{eqnarray}
would yield

\begin{eqnarray}
\mathcal{B}_{ii} &=&\frac{\sin ^{2}\gamma }{C_{ii}e^{-p_{i}}-B_{ii}e^{p_{i}}}%
e^{-p_{i}}\Big(-2b_{i}b_{i}^{^{\prime }}\sum_{j\neq i}^{n}\Big(\frac{\cos
(x_{i}+x_{j})}{\sin (x_{i}+x_{j}+\gamma )\sin (x_{i}+x_{j}-\gamma )}
\nonumber \\
&&+\frac{\cos (x_{i}-x_{j})}{\sin (x_{i}-x_{j}+\gamma )\sin
(x_{i}-x_{j}-\gamma )}\Big)  \nonumber \\
&&+\sum_{j\neq i}^{n}\Big(\frac{b_{i}b_{j}^{^{\prime }}\cot (x_{ik})}{\sin
^{2}(x_{ik}-\gamma )}+\frac{b_{i}^{^{\prime }}b_{j}\cot (x_{ik})}{\sin
^{2}(x_{ik}+\gamma )}+\frac{b_{i}b_{j}\cot (x_{i}+x_{j})}{\sin
^{2}(x_{i}+x_{j}-\gamma )}+\frac{b_{i}^{^{\prime }}b_{j}^{^{\prime }}\cot
(x_{i}+x_{j})}{\sin ^{2}(x_{i}+x_{j}+\gamma )}\Big)\Big),  \nonumber \\
\mathcal{C}_{ii} &=&e^{2p_{i}}\mathcal{B}_{ii}.
\end{eqnarray}
As for the explicit expression of $\mathcal{B}_{ii},\mathcal{C}_{ii},$ we
have more simple form for small $n:$

\begin{itemize}
\item  for $n=2,$
\begin{eqnarray}
\mathcal{B}_{ii} &=&0,  \nonumber \\
\mathcal{C}_{ii} &=&0,
\end{eqnarray}

\item  for $n=3,$
\begin{eqnarray}
\mathcal{B}_{ii} &=&\frac{2}{v_{i}}e^{-p_{i}}\cos \gamma \cos (2x_{i})\sin
^{3}\gamma ,  \nonumber \\
\mathcal{C}_{ii} &=&\frac{2}{v_{i}}e^{p_{i}}\cos \gamma \cos (2x_{i})\sin
^{3}\gamma   \nonumber \\
&=&e^{2p_{i}}\mathcal{B}_{ii},
\end{eqnarray}

\item  for $n=4,$
\begin{eqnarray}
\mathcal{B}_{ii} &=&\frac{2}{v_{i}}e^{-p_{i}}\cos \gamma \cos (2x_{i})\sin
^{3}\gamma \Big(2\cos x_{i}\sin ^{2}\gamma +\sum_{j\neq i}^{n}\sin
(x_{i}+x_{j})\sin (x_{ij})\Big),  \nonumber \\
\mathcal{C}_{ii} &=&\frac{2}{v_{i}}e^{p_{i}}\cos \gamma \cos (2x_{i})\sin
^{3}\gamma \Big(2\cos x_{i}\sin ^{2}\gamma +\sum_{j\neq i}^{n}\sin
(x_{i}+x_{j})\sin (x_{ij})\Big)  \nonumber \\
&=&e^{2p_{i}}\mathcal{B}_{ii}.
\end{eqnarray}
\end{itemize}

We have checked that $L,M$ satisfy the Lax equation of Eq.(\ref{laxequ})
which equivalent to the equations of motion Eq.(\ref{equ1}) and Eq.(\ref
{equ2}) at least for $n=2,3,4,5,6$ with the help of computer.

\section{Nonrelativistic limit to the Calogero-Moser system}

\setcounter{equation}{0} \label{limit}

It is natural that we must verify if the nonrelativistic limit is correct.
The procedure can be achieved by rescaling $p_{i}\longmapsto \beta p_{i}$, $%
\gamma \longmapsto \beta \gamma $ while letting $\beta \longmapsto 0^{+}$%
(here, $0^{+}$ is to avoid undefinable limit of $\mathcal{\ B } _{ii}$ and $%
\mathcal{C}_{ii}$ when $n=2$)$,$ and making a canonical transformation

\begin{equation}
p_{i}\longmapsto p_{i}+\gamma \left( \sum_{k\neq i}^{n}\left( \cot
(x_{ik})+\cot (x_{i}+x_{k})\right) \right) ,
\end{equation}
such that
\begin{eqnarray}
L &\longmapsto &Id+\beta L_{CM}+O(\beta ^{2}), \\
M &\longmapsto &2\beta M_{CM}+O(\beta ^{2}),
\end{eqnarray}
and

\begin{equation}
H\longmapsto 2n+2\beta ^{2}H_{CM}+O(\beta ^{2}).
\end{equation}
$L_{CM}$ can be expressed as
\begin{equation}
L_{CM}=\left(
\begin{array}{ll}
A_{CM} & B_{CM} \\
-B_{CM} & -A_{CM}
\end{array}
\right) ,
\end{equation}
where
\begin{eqnarray}
(A_{CM})_{ij} &=&\delta _{ij}p_{i}+(1-\delta _{ij})\frac{\gamma }{\sin
(x_{ij})},  \nonumber \\
(B_{CM})_{ij} &=&(1-\delta _{ij})\frac{\gamma }{\sin (x_{i}+x_{j})}.
\end{eqnarray}
$M_{CM}$ is
\begin{equation}
M_{CM}=\left(
\begin{array}{ll}
\mathcal{A}_{CM} & \mathcal{B}_{CM} \\
\mathcal{B}_{CM} & \mathcal{A}_{CM}
\end{array}
\right) ,
\end{equation}
where
\begin{eqnarray}
(\mathcal{A}_{CM})_{ij} &=&-\delta _{ij}\sum_{k\neq i}^{n}\Big(\frac{\gamma
}{\sin ^{2}x_{ik}}+\frac{\gamma }{\sin ^{2}(x_{i}+x_{k})}\Big)+(1-\delta
_{ij})\frac{\gamma \cos (x_{ij})}{\sin ^{2}x_{ij}},  \nonumber \\
(\mathcal{B}_{CM})_{ij} &=&(1-\delta _{ij})\frac{\gamma \cos (x_{i}+x_{j})}{%
\sin ^{2}(x_{i}+x_{j})},  \label{mcm1}
\end{eqnarray}
which coincides with the form given in \cite{op,bcs} with the difference of
a constant diagonalized matrix.

The Hamiltonian of the $D_{n}$-type CM model can be given by
\begin{eqnarray}
H_{CM} &=&\frac{1}{2}\sum_{k=1}^{n}p_{k}^{2}-\gamma ^{2}\sum_{k<i}^{n}\Big(%
\frac{1}{\sin ^{2}x_{ik}}+\frac{1}{\sin ^{2}(x_{i}+x_{k})}\Big)  \nonumber \\
&=&\frac{1}{4}trL^{2}.
\end{eqnarray}
The $L_{CM}$, $M_{CM}$ satisfy the Lax equation

\begin{eqnarray}
\dot{L}_{CM}=\{L_{CM},H_{CM}\}=\lbrack M_{CM},L_{CM}\rbrack.
\end{eqnarray}

\noindent

\textbf{Remarks:}

\begin{quotation}
As far as the forms of the Lax pair for the rational-type RS and CM systems
are concerned, we can get them by making the following substitutions
\begin{eqnarray}
\sin x &\rightarrow &x,  \nonumber \\
\cos x &\rightarrow &1,
\end{eqnarray}
for all of the above statements.
\end{quotation}

\section{Summary and discussions}

\label{sum}

In this paper, we have presented the Lax pair for the classical $n-$particle
trigonometric $D_{n}$ Ruijsenaars-Schneider model together with its rational
limit. We give one explicit form of the Lax pair for small $n$ such as $2,3,4
$ and show the involutive Hamiltonians could be generated by the
corresponding Lax matrix. For generic $n$ we have constructed the Lax pair
and given a general form for it though lacking of a complete proof. But its
correctness could be checked at least for $2\leq n\leq 6$. In the
nonrelativistic limit, this system naturally leads to well known
Calogero-Moser system associated with the root system of $D_{n}$.

Actually, our original aim is to expand our constructions to
the dynamical systems associated with all of the root systems.
As suggested in \cite{GN95} and \cite{aru}, $A_{n-1}$ RS
model appeared in the Hamiltonian reduction procedure
applied to the cotangent bundle over centrally extended
current group while the cotangent bundle over the
centrally extended current algebra was used to obtain the
elliptic Calogero-Moser model\cite{GN94,AM95}. It is natural
to expect similar results to other root systems.
Unfortunately, we fail in the corresponding constructions for the
systems associated with the root systems other than
$A_{n-1}$. In fact, as was analyzed in \cite{hm}, there are several
obstructions to extend the constructions.
Alternatively, they used the so-called ``structure group'',
which related to  Weyl reflections, to process symplectic
reduction to construct the CM systems associate with Hitchin
system where the embedding was not even a group but a semi-direct
product of groups. Moreover, one has the $BC_n$ CM and RS
systems but they do not even correspond to groups. So the
more general and elegant method to universal construction
for the RS systems must combine all of characters appeared
in previous results and get over the obstructions mentioned
above.

On the other hand, a more concrete method is to use
pure algebraic construction, which has made great success
for CM systems\cite{hp1,bcs,bcs2,bcs3,bcs1}. In the present
paper, we try to following this idea and work out partial
result for $D_{n}$ RS system where some formulas such as
Eq.(\ref{reflection}), (\ref{lmat2}) have revealed some characters
of Weyl reflections. Though we haven't obtained universal description of this system,
we hope these results would reveal some essential
ingredient for its integrability and shed some light on
universal characters for generic RS systems. At the same time, we
address it an interesting aspect that the reduction procedure of using ``structure group" corresponding to
RS systems and fixing certain momentum map
suggested in \cite{N96,hm} may be a potential method to accomplish
the complete generalization for RS systems associated with all of
simple Lie algebra and even to all of root systems.
Moreover, the issue for getting the $r$-matrix structure for
this model is deserved due to the success of calculation for the
trigonometric $BC_{n}$ RS system by Avan \textit{et al} in \cite{Avan}.

\section*{Acknowledgement}

One of the authors K. Chen is grateful to professors K.J. Shi, L. Zhao and
W.L. Yang for their encouragement and valuable discussions.


\vspace{1pt}

\vspace{1pt}

\begin{thebibliography}{99}
\bibitem{r3}  Ruijsenaars S N M and Schneider H 1986 \textit{Ann. Phys.}
\textbf{170} 370

\bibitem{r1}  Ruijsenaars S N M 1987 \textit{Commun. Math. Phys.} \textbf{%
110 } 191

\bibitem{r2}  Ruijsenaars S N M 1988 \textit{Commun. Math. Phys.} \textbf{%
115 } 127

\bibitem{h1}  Hasegawa K 1997 \textit{Commun. Math. Phys.} \textbf{187} 289

\bibitem{nksr}  Nijhoff F W, Kuznetsov V B, Sklyanin E K and Ragnisco O 1996
\textit{J. Phys.} \textbf{A}: \textit{Math. Gen.} \textbf{29}, L333

\bibitem{n}  Nekrasov N 1998 \textit{Nucl. Phys.} \textbf{B531} 323

\bibitem{bm1}  Braden H W, Marshakov A, Mironov A and Morozov A 1999
\textit{Nucl. Phys. }\textbf{B558} 371

\bibitem{hp2}  D'Hoker E and Phong D H 1999 Lectures on supersymmetric
Yang-Mills theory and integrable Systems \textit{Preprint }\texttt{\
hep-th/9912271}

\bibitem{Marbook}  Marshakov A 1998 \textit{Seiberg-Witten Theory and
Integrable Systems} (Singapore: World Scientific)

\bibitem{g00}  Gorsky A and Mironov A 2000 Integrable many-body systems and
gauge theories \textit{Preprint} \texttt{hep-th/0011197}

\bibitem{op}  Olshanetsky M A and Perelomov A M 1981 \textit{Phys. Rep.}
\textbf{71} 314; Perelomov A M 1990 \textit{Integrable Systems of Classical
Mechanics and Lie Algebras} (Boston, MA: Birkh\"{a}user)

\bibitem{in}  Inozemtsev V I 1989 \textit{Lett. Math. Phys.} \textbf{17} 11

\bibitem{hp1}  D'Hoker E and Phong D H 1998 \textit{Nucl. Phys.} \textbf{%
B530 } 537

\bibitem{bcs}  Bordner A J, Corrigan E and Sasaki R 1998 \textit{Prog.
Theor. Phys. }\textbf{100} 1107

\bibitem{bcs2}  Bordner A J, Sasaki R and Takasaki K 1999 \textit{Prog.
Theor. Phys.} \textbf{101} 487

\bibitem{bcs3}  Bordner A J and Sasaki R 1999 \textit{Prog. Theor. Phys.}
\textbf{101} 799

\bibitem{bcs1}  Bordner A J, Corrigan E and Sasaki R 1999 \textit{Prog.
Theor. Phys. }\textbf{102} 499.

\bibitem{hm}  Hurtubise J C and Markman E 1999 Calogero-Moser systems and
Hitchin systems \textit{Preprint} \texttt{math/9912161}.

\bibitem{bms}  Bordner A J, Manton N S and Sasaki R 2000 \textit{Prog.
Theor. Phys. }\textbf{103} 463

\bibitem{kps}  Khastgir S P, Pocklington A J and Sasaki R 2000 \textit{J.
Phys.} \textbf{A}: \textit{Math. Gen. }33 9033

\bibitem{op1}  Olshanetsky M A and Perelomov A M 1983 \textit{Phys. Rep.}
\textbf{94} 313

\bibitem{bc}  Bruschi M and Calogero F 1987 \textit{Commun. Math. Phys.}
\textbf{109} 481.

\bibitem{kz}  Krichever I and Zabrodin A 1995 \textit{Usp. Math. Nauk}
\textbf{50:6} 3

\bibitem{s1}  Suris Y B 1996 Why are the rational and hyperbolic
Ruijsenaars-Schneider hierarchies governed by the same $R$-operators as the
Calogero-Moser ones? \textit{Preprint }\texttt{hep-th/9602160}

\bibitem{s2}  Suris Y B 1997 \textit{Phys. Lett.} \textbf{A225} 253

\bibitem{aru}  Arutyunov G E, Frolov S A and Medvedev P B 1997 \textit{J.
Math. Phys.} \textbf{38} 5682

\bibitem{ko1}  Komori Y and Hikami K 1998 \textit{J. Math. Phys.} \textbf{39}
6175

\bibitem{di}  van Diejen J F 1994 \textit{J. Math. Phys. }\textbf{35} 2983

\bibitem{di1}  van Diejen J F 1995 \textit{Compositio. Math.} \textbf{95} 183

\bibitem{h2}  Hasegawa K, Ikeda T and Kikuchi T \textit{J. Math. Phys.}
\textbf{40} 4549

\bibitem{kai3}  Chen K, Hou B Y and Yang W L 2001 \textit{Chin. Phys.} \textbf{10} 550

\bibitem{Che00}  Chen K, Hou B Y and Yang W L 2000 \textit{J. Math. Phys.}
\textbf{41} 8132

\bibitem{Avan}  Avan J and Rollet G 2000 $BC_{n}$ Ruijsenaars-Schneider
models: R-matrix structure and Hamiltonians \textit{Preprint}
\texttt{hep-th/0008174}

\bibitem{Che01}  Chen K, Hou B Y and Yang W L 2000 \textit{J. Math. Phys.}
\textbf{42} (2001) 4894



\bibitem{GN95}  Gorsky A and Nekrasov N 1995 \textit{Nucl. Phys.} \textbf{B436}
582

\bibitem{GN94} Gorsky A and Nekrasov N 1994 Elliptic Calogero-Moser
system from two-dimensional current algebra \textit{Preprint}
 \texttt{hep-th/9401021}

\bibitem{AM95} Arutyunov G E and Medvedev P B 1995
Generating equation for $r$-matrices related to dynamical
systems of Calogero type; \textit{Preprint}
\texttt{hep-th/9511070}

\bibitem{N96}  Nekrasov N 1996 \textit{Commun. Math. Phys.} \textbf{180}
587

\end{thebibliography}
\end{document}